# LARGE SCALE PECULIAR VELOCITIES: EFFECTS FROM SUPERCLUSTERS

Roberto Scaramella[1]

[1]*Osservatorio Astronomico di Roma, 00040 Monteporzio Catone, Italy.*

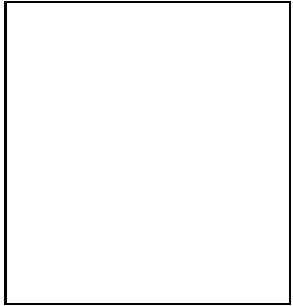

**Abstract**

We study the gravitational influence of very large scale structures, as traced by clusters of galaxies, on the Local Group [LG] motion and the large scale flows.

We derive from the distribution of Abell clusters within 300 Mpc/h the overdensity field on a 3–D grid of spacing 5 Mpc/h, then we solve the Poisson equation for the peculiar potential and finally obtain the peculiar velocity field.

Quite interestingly, from *this global solution* we:

i) recover within $\sim 10^o$ the direction of the LG motion in the Cosmic Microwave Background [CMB] frame,

ii) find that the Great Attractor itself moves wrt to the CMB frame,

iii) derive $\beta_c^{-1} \equiv b_c \Omega_0^{-0.6} = 5.3 \pm 0.20$ from a preliminary comparison with the Mark II catalog of peculiar velocities,

iv) derive estimates for the bulk flow in spheres which fairly agree with the level derived from POTENT and the Spiral samples,

v) find that the Lauer & Postman [L&P] bulk flow has too a large amplitude to be in agreement with our results.

## 1 Sample and method

A fuller discussion can be found in [5]. Our starting point are the Abell and ACO catalogs of clusters. Here we use data from an all–sky sample of $\simeq 860$ clusters of galaxies, obtained by joining the Abell and ACO catalogs, considering a region complete in volume with limiting depth of 300 Mpc/h and with $|b| > 20^o$. We place the clusters on a $128^3$ grid, of side 640 Mpc/h. The clusters are allocated to the nearest grid point two different weights (number and mass, see [5]) which give an estimate of the intrinsic uncertainty of the results. We divide this field (Gaussianly smoothed with $\sigma = 15$ Mpc/h) by the average of 200 pseudo–random samples [6] and solve the Poisson equation for the peculiar potential of this field with periodic b.c. Finally we derive the linear velocity field by differencing the potential.

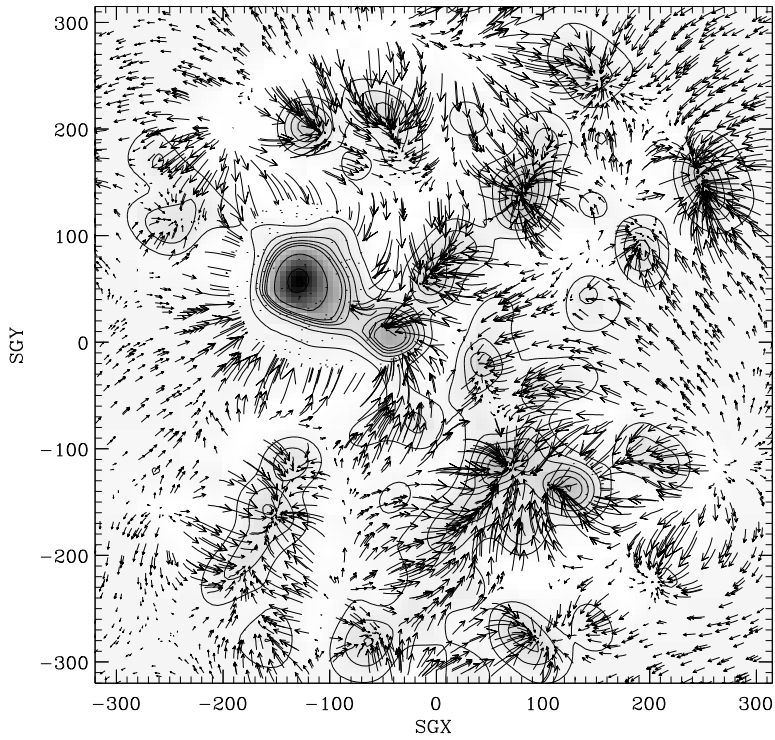

Figure 1: *We show the overdensity field and peculiar motions in the supergalactic plane. Units are in Mpc/h.*

## 2 Qualitative results

In Fig. 1 we show $\delta_c(\vec{x})$ for a slice of 15 Mpc/h half–width centered on the SG plane in order to show the effect of this procedure, together with the derived velocity field in the plane. See [5] for a list of technical problems, assumptions, checks and refinements to be possibly dealt with, as well as a larger number of figures of the field and flows.

From the flows we derive the picture of a complex pattern, in which a few qualitative conclusions can be drawn (see [5]). Among these, the prediction of an offset of the density peak with the minumum of the peculiar velocity field at the GA position is in a remarkable agreement with the latest results from POTENT applied to the Mark III catalog [2]. It must be noticed that if we force the overdensity field to be zero beyond a depth of 100 Mpc/h, that is with no influence from very large scales, this offset disappears.

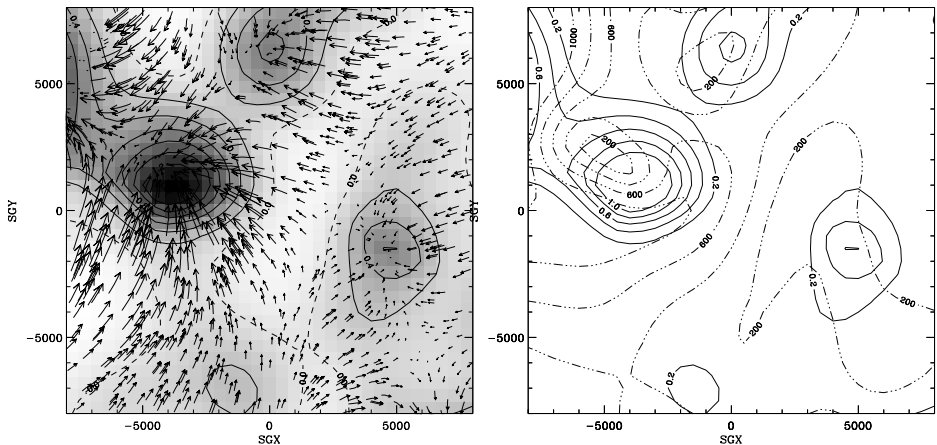

Figure 2: *In the left panel we show an enlargement of the inner region of Fig.2, the one presently seen by catalogs of galaxies. In the right panel we show density isocontours superimposed to peculiar velocities isocountours. Notice the offset at the GA position.*

## 3 Estimates of the bias factor

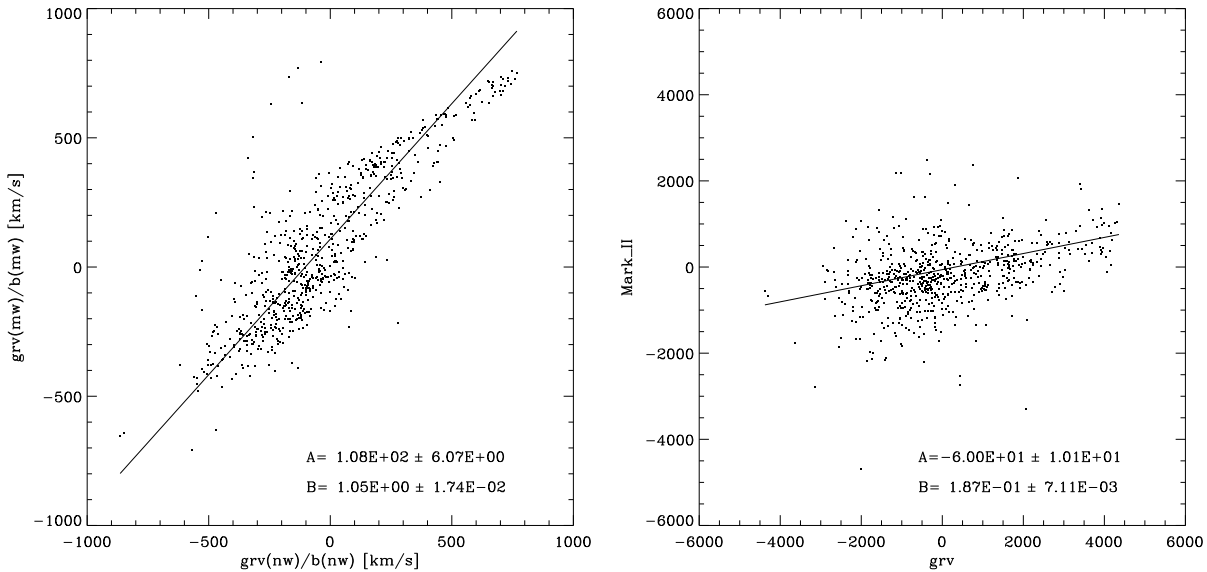

Figure 3: *In the left panel we show the dispersion of the values predicted according to the two different weights at the location of the Mark II galaxies. In the right panel we show the fit for the value of $\beta_c$ from the average of predictions from the left panel against the Mark II values.*

Now, under the strong hypotheses of: i) validity of linear theory, and ii) simple proportionality of the $\delta_c(\vec{x})$ field and the true mass fluctuation field, $\delta_\rho(\vec{x})$, one would be able to derive the 3D peculiar flows on quite large scales, and in particular the one of the LG. We recover the latter within $\approx 10^o$ [5].

From normalization of the amplitude one can get estimates for $\beta_c$, but a better way is likely to be an estimate which is based not on a single point (cf [1]), but on comparison with one of the catalogs of peculiar veocities. At this stage we make a preliminary comparison with the galaxies of the Mark II catalog grouped as in [7].

From a preliminary fit with errors in both variables, we obtain (right panel of Fig. 3) an estimate of $\beta_c = 0.19 \pm 0.01$, which translates into $b_c \Omega_0^{-0.6} = 5.3 \pm 0.20$. Therefore, if one rescales by assuming reasonable but uncertain values for relative biases ($b_{c,o} \equiv 3.5$, $b_{c,I} \equiv 4.5$ [4]), one obtains the estimates of $\Omega_0 \approx 0.5$ if $b_o \equiv 1$, or $\Omega_0 \approx 0.8$ if $b_I \equiv 1$, with uncertainties much larger than the formal ones.

## 4 Not too large bulk flows on large scales

Even though it furnishes a single point (to the large experimental uncertainties adds the one due to cosmic variance), the motion of a sphere centered on us as a whole wrt the CMB frame is an interesting value, with controversial results which date back to the Rubin–Ford effect. In Fig. 4 we show our preliminary solutions, together with quite recent determinations (the spiral sample results have been presented by Da Costa at this conference). We see that the amplitude of the motion we derive is in good agreement with that from the galaxy samples, while differs from the one derived from clusters [3] (for a depth of 100Mpc/h L&P find $700 \pm 180$ km/s vs our $250 \pm 50$ km/s). One must bear in mind, though, problems due to boundary conditions and unsurveyed areas in our model.

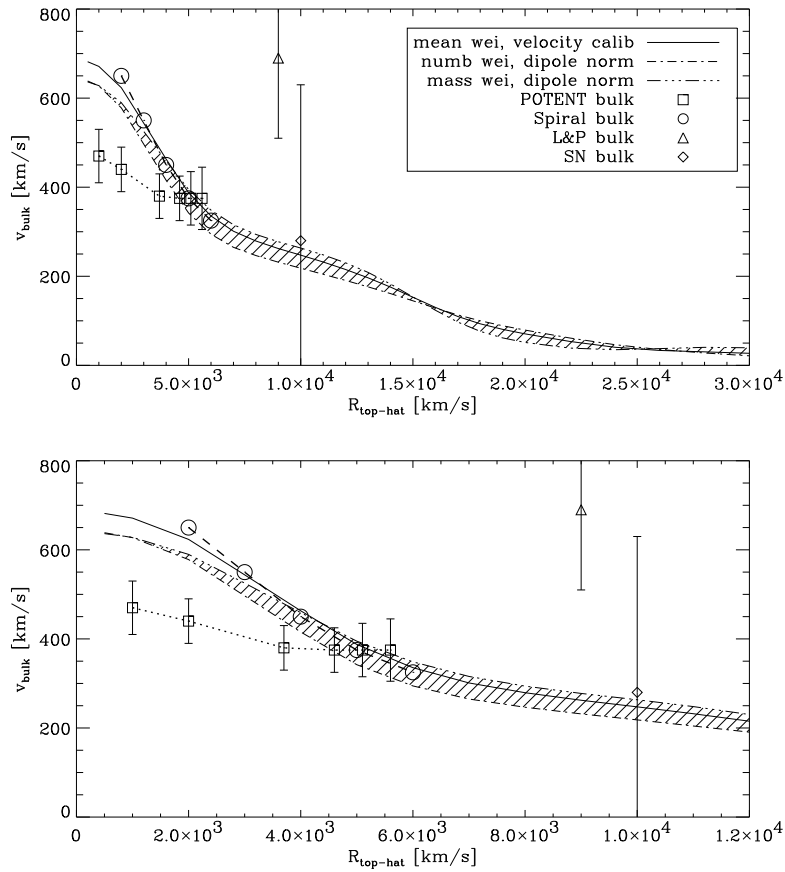

Figure 4: *We show our solutions for the amplitude of the bulk motion in spheres centered on the LG.*